# How Does Science Education Research Respond to Sociopolitical Change? A BERTopic Analysis of Korean Research

Jibeom Seo[1], Junghyo Jo[1,2,3], Sonya N. Martin[1] and Taejin Byun[4,5,*]

[1]Department of Science Education, Seoul National University, Seoul, Republic of Korea.

[2]Center for Theoretical Physics and Artificial Intelligence Institute, Seoul National University, Seoul, Republic of Korea.

[3]School of Computational Sciences, Korea Institute for Advanced Study, Seoul, Republic of Korea.

[4*]Department of Science Education, Gwangju National University of Education, Gwangju, Republic of Korea.

[5*]Department of Teaching and Learning, Bayh College of Education, Indiana State University, Terre Haute, Indiana, United States.

**Abstract**

Research fields do not evolve in isolation: their questions and priorities shift with policy, curriculum reform, and broader social change. Analyzing published literature can reveal not only how a field matures but also how it responds to these conditions. Prior work in science education has focused on identifying research topics and their trends, but paid less attention to the external conditions in which research is produced. We examine Korean science education research from 2008 to 2025, a case in which centralized curriculum revision, government education initiatives, and demographic decline are prominent. Using BERTopic, an embedding-based topic modeling technique, we identify major topics and temporal trends, and analyze their associations with selected sociopolitical factors. We interpret each topic and distinguish three groups: sociopolitical, subject-specific, and student-related topics. Within the first group, science teacher professionalism and curriculum implementation, science education for gifted students, and STEAM education show the strongest associations with sociopolitical conditions, such as government policy initiatives and declining enrollment in science-gifted education, whereas digital-based science education does not. The subject-specific and student-related groups, by contrast, show no comparable movement and are not linked to the external indicators we examine; this pattern is interpreted as reflecting stronger disciplinary grounding. Taken together, these patterns suggest that a topic's anchoring to policy and practice or to academic disciplines shapes how closely it tracks external change. This helps explain why some research agendas move with their national context while others hold steady, and why the same topic may develop differently across countries.

## Introduction

Large-scale literature reviews have shown how science education research changes over time by identifying research topics and trends (Chang & Na, 2022; Lin et al., 2025; Odden et al., 2021). Much of this work maps what topics emerge and how they rise or fall, and there is room to build on it in two directions: linking topic trends more systematically to the external conditions that accompany them, and interpreting why different topics follow different trajectories. Because prior studies have often drawn on international journals (Kahraman, 2026; Lin et al., 2025; Odden et al., 2021), less is known about how research agendas take shape within national education systems, where local conditions can influence the formation and persistence of research topics (Faisal & Martin, 2019; Wang et al., 2025). This study aims to fill this research gap by examining papers from three Korean[1] science education journals and connecting the identified topics to sociopolitical conditions.

Korea offers a particularly instructive case. Over the period examined here, science education has been closely tied to a fast-moving policy environment: successive national curriculum revisions (Choi & Choi, 2016), STEAM initiatives (N.-H. Kang, 2019), and a sharp demographic decline in the school-age population (Jung, 2024). This makes the relationships between national policy, social change, and research agendas unusually visible. Because policy is set centrally and reaches schools quickly, Korea clearly shows a dynamic that many centralized education systems share: research agendas can shift in step with sociopolitical change. Examining this case can therefore help researchers in other national contexts see how their own systems may respond to similar pressures. Within this setting, science education research includes topics grounded in established academic contexts (Tsai & Wen, 2005), as well as topics related to contexts of application, such as government initiatives (Odden et al., 2021). The former reflect long-standing concerns shared across science education communities, while the latter respond more directly to practical and policy-driven demands. The Korean case allows these different patterns to be examined within the same national research system. In this study, we distinguish between topics with distinctive temporal trends and those without. We investigate whether topics with distinctive trends are related to sociopolitical conditions, including national curriculum revisions, government policies, and demographic change. This study also examines whether topics without distinctive trends have strong disciplinary characteristics. This comparison clarifies how different types of research topics develop within a national research system.

[1] In this paper, “Korea” refers to the Republic of Korea (South Korea).

Based on this rationale, this study examines science education research published in three Korean journals from 2008 to 2025. We use BERTopic, an embedding-based topic modeling method for identifying topics in a large text dataset (Grootendorst, 2022), to identify major research topics and trace their temporal trends. We then compare selected topic trends with external indicators related to curriculum revision periods, government initiatives, and demographic change. Our central argument is that a topic's responsiveness to external change depends on how it is anchored: application-oriented topics, closely tied to policy and funding, shift with these conditions, whereas topics supported by established academic communities remain comparatively stable. By combining topic modeling with contextual data, this study provides empirical evidence for this argument and shows how national literature analysis can complement studies centered on international journals. The following research questions guide this study:

- RQ1. What major topics and temporal trends can be identified in Korean science education research from 2008 to 2025?
- RQ2. How are temporal changes in Korean science education research topics associated with sociopolitical conditions?
- RQ3. What characteristics distinguish research topics that show stronger responsiveness to external conditions from those that do not?

## Literature Review

### Overview of Korean science education

This subsection provides an overview of science education in Korea to establish the necessary background for this study.

#### *Historical Background and Curriculum System*

In the aftermath of the Korean War (1950–1953), Korea faced the urgent task of training a skilled STEM workforce for national reconstruction and development (Leem & Kim, 2013). This priority was reflected in the establishment of the centralized national curriculum, through which national policy directions have exerted a decisive influence on Korean science education (Choi & Choi, 2016; Leem & Kim, 2013). Korea issued its first curriculum in 1954, and revised curricula have been regularly released since then; the *2022 Revised National Curriculum* was announced as the 11th version. Revised curricula have adjusted subjects and their content to reflect emerging needs and circumstances. However, implementation in schools typically lags behind the announcement by a few years.

Under this curriculum system, school science is organized into two distinct phases[2] (Ministry of Education, 2015, 2022). First, while science is integrated with other subjects in Grades 1–2, it is taught as a single, mandatory core subject in Grades 3–9 to foster the foundational scientific literacy necessary for high school science. Second, at the high school level, which covers Grades 10–12, the national curriculum transitions to a more specialized, credit-based system tailored to students' individual career paths. Specifically, for Grades 11–12, there are general elective and career elective courses across four disciplines: physics, chemistry, biology, and earth science. Consistent with the discipline-based curriculum, secondary teachers are prepared in physics, chemistry, biology, or earth science teacher education programs, whereas elementary school teachers are prepared to teach all subjects (N.-H. Kang, 2019).

#### *Educational Practices and Challenges*

Korean students consistently rank among the world's top performers in science achievement (Mullis et al., 2020), yet their affective attitudes toward science remain comparatively low. Despite strong problem-solving skills, they report low confidence in science and little enjoyment in learning science (Geesa et al., 2019; Yoon et al., 2014). These attitudes can be partly attributed to prevailing pedagogical practices. In Korea, science education has predominantly centered on knowledge-based rather than process-focused assessment (Park et al., 2016). Furthermore, due to intense competition for college admissions, similar to China, Japan, and Singapore (Kennedy, 2007), students tend to focus on test-taking strategies rather than authentic inquiry (Kwon et al., 2017). To raise affective attitudes toward science and encourage students to choose STEM careers, the Korean government introduced STEAM (Science, Technology, Engineering, Arts, and Mathematics) education in 2011 (N.-H. Kang, 2019). This policy initiative aims to enhance student interest and understanding in science and technology by integrating the humanities and arts (A) into STEM.

### Overview of Methods in Science Education Literature Research

This subsection reviews three methods commonly used to extract topics or keywords from science education literature: manual content analysis (MCA), semantic network analysis (SNA), and latent Dirichlet allocation (LDA). We then introduce BERTopic, which was used in this study.

#### *MCA, SNA, and LDA*

MCA involves researchers reading and coding papers according to a predefined classification system,

---

[2] Korea's education system follows a 6-3-3 structure with a national curriculum that covers grades 1 to 12: 6 years of primary school, 3 years of lower secondary school (middle school), and 3 years of upper secondary school (high school).

which allows in-depth interpretation of each study and its context (Cavas, 2015; Lin et al., 2025). However, it is time-consuming, labor-intensive, and can be influenced by coder subjectivity. SNA addresses some limitations of MCA by modeling keywords as nodes and their co-occurrences as links, thereby allowing researchers to analyze conceptual relationships and influential keywords across large datasets (Lee & Kim, 2018; Seo et al., 2024; Tang et al., 2024). Yet SNA abstracts keywords from their surrounding context, which can reduce semantic nuance. LDA further supports large-scale and replicable literature analysis by automatically grouping frequently co-occurring words into latent topics (Blei et al., 2003; Chang & Na, 2022; Odden et al., 2021). Nevertheless, because LDA relies on bag-of-words representations, it does not fully capture semantic relationships among words (Grootendorst, 2022).

***BERTopic***

BERTopic, a transformer-based topic modeling method (Grootendorst, 2022), can reduce the coding burden required for MCA and support multidimensional analysis of the data. Unlike SNA and LDA, BERTopic uses Sentence-BERT (SBERT) to capture the contextual meaning of texts by transforming them into vector representations (Reimers & Gurevych, 2019). This enables the model to cluster documents by semantic similarity rather than lexical co-occurrence. Figure 1 summarizes the main procedure of BERTopic. First, documents are converted into vector representations using a pre-trained language model, such as SBERT. Second, these vectors are grouped into semantic clusters with similar meanings (e.g., smiley and sad faces forming distinct groups). Lastly, topic representations are extracted from each cluster using c-TF-IDF, which weights words by how strongly they characterize one cluster relative to the others.

**Fig. 1**

*Conceptual diagram of the BERTopic process*

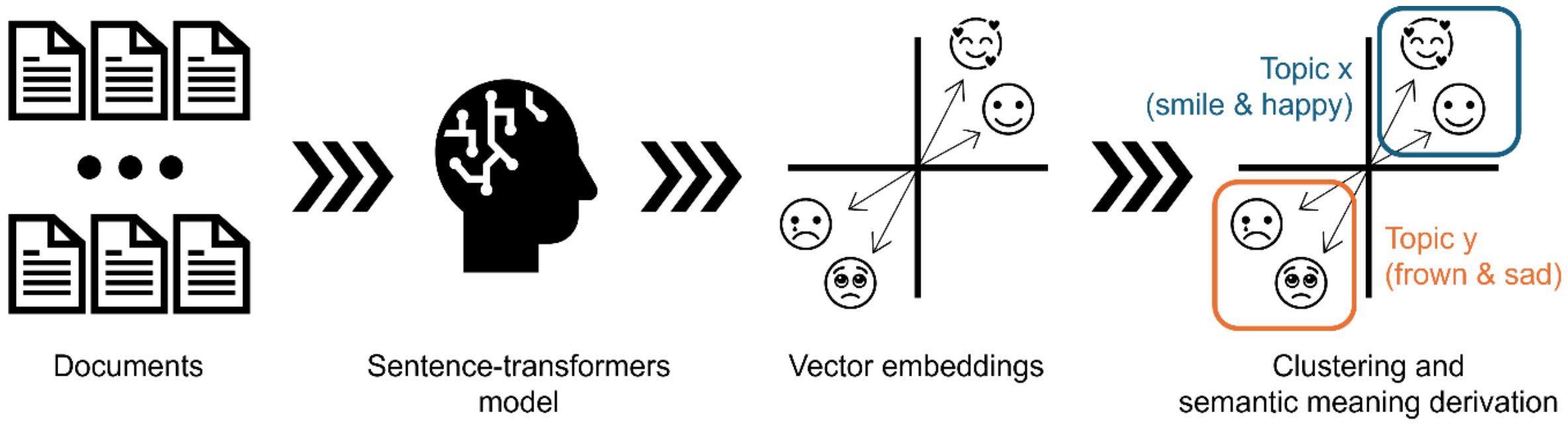

# Methodology

## Data Collection

We established criteria for selecting target journals to analyze science education research in Korea. First, we chose peer-reviewed journals specializing in science education with a publication history of at least 15 years. Second, to obtain sufficient data, we selected journals published at least three times a year. Third, we excluded journals limited to specific subfields (e.g., physics, gifted education) or those limited to particular school levels. Consequently, three journals were selected that met these criteria: *Journal of the Korean Association for Science Education* (JKASE), *Journal of Science Education* (JSE), and *School Science Journal* (SSJ). Additionally, since SSJ publication records are available from 2008, we included all papers published from that year onward.

In total, we collected 2,246 papers. Figure 2 presents the annual number of publications, both in total and by journal. To identify topics and their trends, we used the publication year and English abstract. In addition, to examine organizational heterogeneity, we analyzed the affiliations of all authors by institution type. Affiliations were grouped into eight categories: K-12 schools, universities and colleges, educational administrative agencies, educational research institutes, other research institutes, science and natural history museums, industry, and others. Each author was assigned one affiliation. Authors affiliated with both “universities and colleges” and “K-12 schools” were assigned to the latter, as these cases usually involved teachers enrolled in graduate programs.

**Fig. 2**

*Annual number of publications in three science education journals from 2008 to 2025*

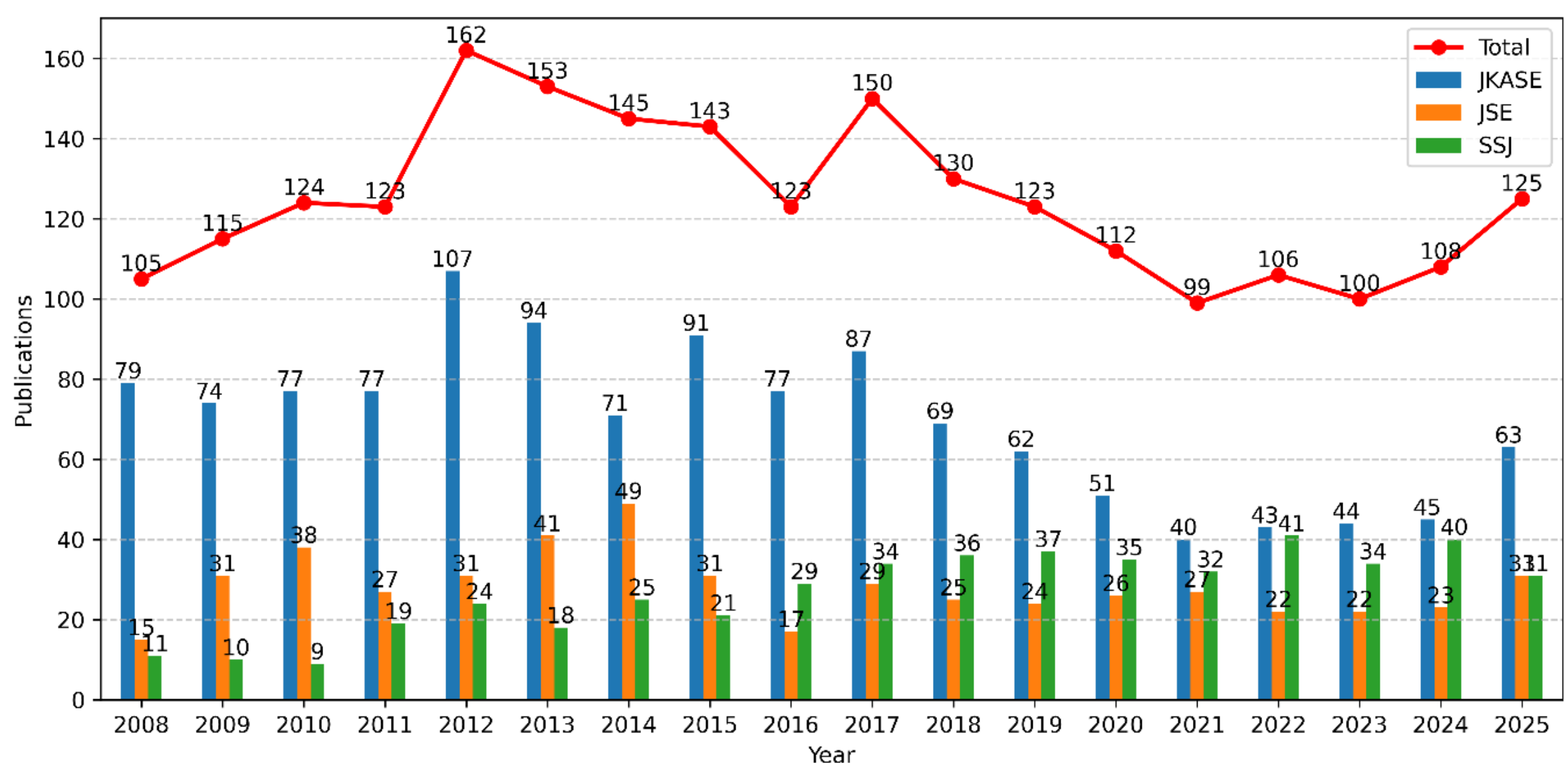

**Model Selection**

We generated 90 candidate models via a grid search over selected hyperparameters of BERTopic. The detailed generation process is described in Supplementary Information 1. From the candidates, we selected well-performing models based on five criteria to ensure both statistical and interpretive rigor: topic frequency (TF), topic coherence (TC), topic diversity (TD), proportion of outliers (PO), and face validity.

- TF is the frequency of the number of topics across all models. A high frequency for a certain number suggests that the dataset likely contains that number of topics inherently.
- TC measures the semantic consistency of words within a topic (Newman et al., 2010); the measure ranges from -1 to 1, and a higher TC indicates that the topic is more interpretable for humans.
- TD measures the proportion of unique words among all top-10 topic words, aggregated across topics; the measure ranges from 0 to 1, where a higher score indicates less redundancy (Dieng et al., 2020).
- PO is the percentage of papers not assigned to any specific topic. These unassigned papers typically address minor or mixed themes, insufficient to constitute a distinct topic. We prioritized models with lower PO to maximize data utilization.
- Face validity is assessed through a qualitative evaluation to ensure that topics are plausible and meaningful within the context of our domain knowledge, complementing the quantitative metrics. This final check ensures that the chosen model is semantically meaningful.

We observed that models with 11 topics appeared most frequently, as shown in Fig. 3. From this 11-topic group, we identified candidate models with above-average TC and TD values, and then narrowed the selection to those with low PO values (less than 40%). The semantic results across these models were nearly identical, with minor differences only in the boundaries of a few topics, and the topics later linked to external conditions appeared consistently; this suggests that the 11-topic solution is stable across hyperparameter settings. Furthermore, in terms of face validity, the topics from this group were also plausible and meaningful in the context of science education research. We therefore selected our final model from the 11-topic group, the one with an optimal balance of low PO and high TC and TD.

However, approximately one third of the papers remained outliers and were excluded from the analysis. To address this, we applied outlier reduction to the chosen model using two document representations: c-TF-IDF vectors and SBERT embeddings. Both assign an outlier paper to the most similar topic based on cosine similarity; the difference between the two appears only in the space where similarity is calculated. In the c-TF-IDF space, a paper is compared with each topic in terms of the words that distinguish that topic from the others, since a high c-TF-IDF score indicates a word that occurs frequently within a given topic but rarely elsewhere. In the embedding space, the paper is compared with the mean embedding of the documents in each topic, and thus reflects the overall semantic context of the abstract. We used both representations so that each assignment is supported by distinctive words and by semantic context, and reassigned only the papers that the two mapped to the same topic, keeping the rest as outliers. This procedure assigned 390 additional papers to topics without degrading the topic representations: face validity was unchanged, and both TC and TD were slightly higher afterward (TC: .410 → .460; TD: .982 → .991).

**Fig. 3**

*Boxplots of (a) topic frequency, (b) topic coherence, (c) topic diversity, and (d) proportion of outliers by number of topics*

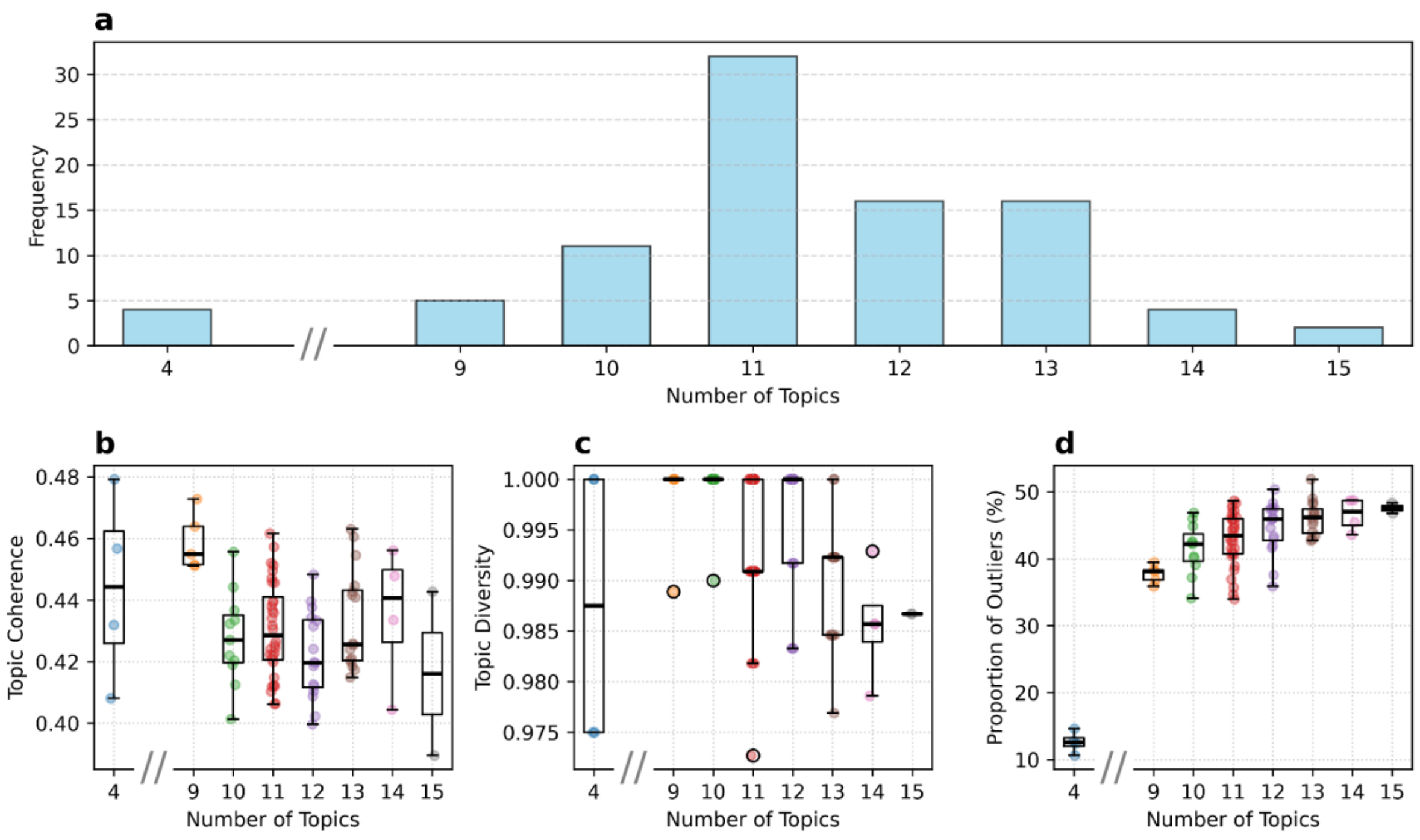

## Topic Interpretation

Lastly, we named the eleven topics derived from BERTopic and interpreted each one based on two sources: its top ten representative words[3] and its top ten representative papers per topic, 110 in total. Using these, we described what kind of concerns each topic reflects, for example, whether it centered on policy, curriculum, and school-level practice, or subject content. We employed investigator triangulation to strengthen the reliability and validity of our topic interpretations (Patton, 2014). Interpretation by a single researcher may lead to over-interpretation, as explanations of topics can be influenced by the researcher's prior knowledge or preconceptions. To reduce this risk, we, as science education experts, repeatedly reviewed and discussed the results because automated techniques such as topic modeling cannot replace the insights of researchers (Asmussen & Møller, 2019).

---

[3] Strictly speaking, "words" refer to N-grams, but we use the term for brevity.

## Results and Discussion

### Topics and Their Interpretations

Using BERTopic, we extracted 11 latent topics from the selected Korean science education publications, presented in Table 1. "Topic −1" consists of outlier documents that were not assigned to any specific topic. The other topics were labeled based on their top ten representative words (see Fig. 4) and papers (see Supplementary Information 2) to provide an interpretable summary. These topics can be grouped into three categories: (1) sociopolitical, (2) subject-specific, and (3) student-related topics. Note that these groups are not mutually exclusive. Since topics often encompass multiple interests, they could reasonably be assigned to more than one group, and topics within the same group do not always exhibit the same characteristics. We grouped each topic by its most prominent orientation, based on its representative words and papers.

**Table 1**

*Topic Information: Number of publications (N* = 2,246*) and name*

| Topic number | Count (Proportion) | Name |
|---|---|---|
| −1 | 373 (16.6%) | (Outlier where papers were not assigned to any specific topic) |
| 0 | 786 (35.0%) | Science Teacher Professionalism and Curriculum Implementation |
| 1 | 227 (10.1%) | Physics and Chemistry Education |
| 2 | 176 (7.8%) | Achievement, Self-efficacy, and Motivation in Science |
| 3 | 160 (7.1%) | Scientific Argumentation |
| 4 | 117 (5.2%) | Science Education for Gifted Students |
| 5 | 80 (3.6%) | Astronomy Education |
| 6 | 77 (3.4%) | Biology Education |
| 7 | 73 (3.3%) | Digital-Based Science Education |
| 8 | 65 (2.9%) | STEAM Education |
| 9 | 56 (2.5%) | Creativity and Thinking in Science |
| 10 | 56 (2.5%) | Ecology Education |

*Note.* Counts before outlier reduction were 763 for outliers and 596, 202, 127, 127, 89, 67, 65, 57, 63, 49, and 41 for Topics 0–10.

**Fig. 4**

*Topic word c-TF-IDF scores*

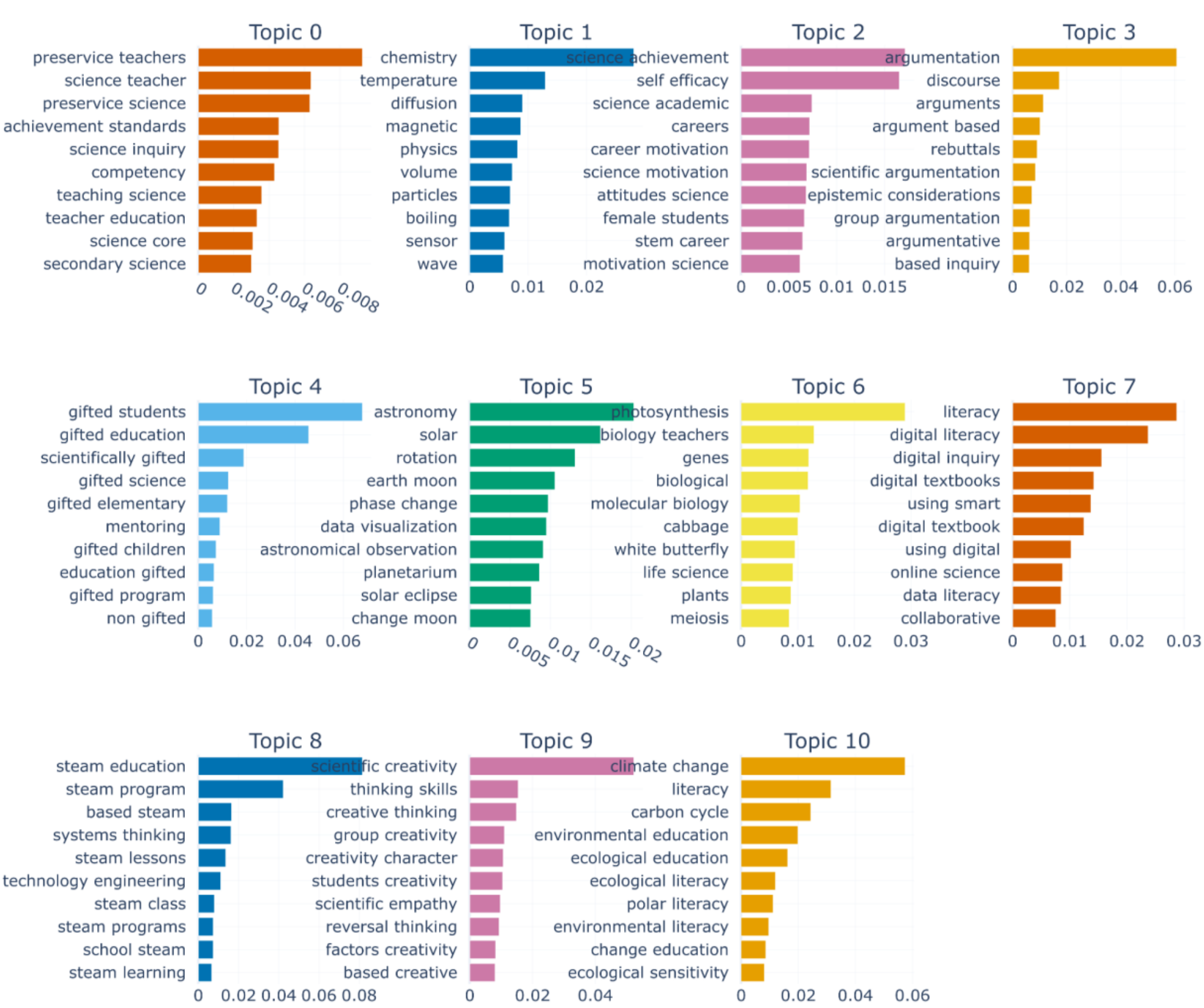

*Note.* Words with higher scores represent the topics more. Topic numbers correspond to those in Table 1.

First, sociopolitical topics are those whose research questions arise from policy, curriculum, and school-level demands. Topic 0 (Science Teacher Professionalism and Curriculum Implementation) covers teacher expertise, teacher education, and the implementation of revised curricula. Topic 4 (Science Education for Gifted Students) focuses on the characteristics of gifted students and teacher development in science gifted education. Topic 7 (Digital-Based Science Education) addresses the use of digital technologies in science classrooms and digital literacy. Topic 8 (STEAM Education) covers the status and development of STEAM education following its introduction. Topic 9 (Creativity and Thinking in Science) examines scientific creativity, thinking, and their relationship; this topic partially overlaps with Topic 4, as some of its papers concern science-gifted students.

Second, subject-specific topics are organized around the content and inquiry traditions of a particular science discipline, which aligns with the subject-based structure of secondary teacher education in Korea. This group includes Topic 1 (Physics and Chemistry Education), Topic 5 (Astronomy Education), Topic 6 (Biology Education), and Topic 10 (Ecology Education). The clustering of physics and chemistry plausibly reflects overlapping content, such as thermodynamics (Seo et al., 2024); the same grouping appeared in an analysis of international journals (Kahraman, 2026), which suggests that this pattern is not specific to the Korean case.

Third, student-related topics take the learner as the object of inquiry: how students think, feel, and reason in science. Their questions come from long-standing research traditions in science education. They include Topic 2 (Achievement, Self-Efficacy, and Motivation in Science) and Topic 3 (Scientific Argumentation). Topic 2 focuses on the relationships among students' affective domains and academic achievement, with some work on STEM career motivation. Topic 3 examines students' engagement in argument-based inquiry.

We also examined the institutional affiliations of authors, presented in Table 2. Across all eleven topics, authors from universities and colleges formed the largest group, ranging from 66.2% (Astronomy Education) to 88.0% (Scientific Argumentation). This dominance is unsurprising for a dataset of peer-reviewed articles, so we note only a few relative differences. Educational research institutes comprise a small share overall but publish more in Topics 0 ($n$ = 101) and 2 ($n$ = 21) than elsewhere. The Korea Institute for Curriculum and Evaluation (KICE) was particularly active in these topics. As a national educational research institute responsible for curriculum design and national examinations in K-12 education, KICE may account for much of this research as part of its public accountability. Topic 5 (Astronomy Education) shows the highest share of K-12 schools (30.8%), reflecting the involvement of teachers. Topic 7 (Digital-Based Science Education) shows the highest share of institutions other than schools and universities (around 10%).

**Table 2**

*Affiliations of all authors by institution type*

| Topic number | K-12 | U & C | EAA | ERI | ORI | S & N | Industry | Others | Total |
|---|---|---|---|---|---|---|---|---|---|
| -1 | 135 (14.0%) | 802 (83.0%) | 5 (0.5%) | 6 (0.6%) | 4 (0.4%) | 1 (0.1%) | 2 (0.2%) | 11 (1.1%) | 966 |
| 0 | 232 (11.5%) | 1,642 (81.4%) | 15 (0.7%) | 101 (5.0%) | 8 (0.4%) | 6 (0.3%) | 0 (0.0%) | 12 (0.6%) | 2,016 |
| 1 | 99 (17.7%) | 445 (79.5%) | 1 (0.2%) | 9 (1.6%) | 2 (0.4%) | 0 (0.0%) | 0 (0.0%) | 4 (0.7%) | 560 |
| 2 | 59 (12.0%) | 399 (81.3%) | 6 (1.2%) | 21 (4.3%) | 3 (0.6%) | 0 (0.0%) | 1 (0.2%) | 2 (0.4%) | 491 |
| 3 | 37 (8.7%) | 375 (88.0%) | 0 (0.0%) | 9 (2.1%) | 2 (0.5%) | 1 (0.2%) | 0 (0.0%) | 2 (0.5%) | 426 |
| 4 | 56 (18.9%) | 234 (79.1%) | 1 (0.3%) | 2 (0.7%) | 2 (0.7%) | 0 (0.0%) | 0 (0.0%) | 1 (0.3%) | 296 |
| 5 | 62 (30.8%) | 133 (66.2%) | 1 (0.5%) | 1 (0.5%) | 2 (1.0%) | 1 (0.5%) | 0 (0.0%) | 1 (0.5%) | 201 |
| 6 | 43 (22.1%) | 151 (77.4%) | 0 (0.0%) | 1 (0.5%) | 0 (0.0%) | 0 (0.0%) | 0 (0.0%) | 0 (0.0%) | 195 |
| 7 | 28 (15.4%) | 136 (74.7%) | 8 (4.4%) | 9 (4.9%) | 0 (0.0%) | 0 (0.0%) | 0 (0.0%) | 1 (0.5%) | 182 |
| 8 | 37 (17.1%) | 171 (79.2%) | 0 (0.0%) | 5 (2.3%) | 0 (0.0%) | 1 (0.5%) | 1 (0.5%) | 1 (0.5%) | 216 |
| 9 | 31 (21.7%) | 108 (75.5%) | 0 (0.0%) | 0 (0.0%) | 0 (0.0%) | 1 (0.7%) | 0 (0.0%) | 3 (2.1%) | 143 |
| 10 | 25 (16.6%) | 117 (77.5%) | 0 (0.0%) | 5 (3.3%) | 0 (0.0%) | 0 (0.0%) | 2 (1.3%) | 2 (1.3%) | 131 |
| Total | 844 (14.4%) | 4,713 (80.7%) | 37 (0.6%) | 169 (2.9%) | 23 (0.4%) | 11 (0.2%) | 6 (0.1%) | 40 (0.7%) | 5,843 |

*Note.* Abbreviations: K-12 schools (K-12), universities and colleges (U & C), educational administrative agencies (EAA), educational research institutes (ERI), other research institutes (ORI), and science and natural history museums (S & N). Topic numbers correspond to those in Table 1.

The visualization of vector embeddings provides an intuitive overview of the data landscape. Figure 5 illustrates how research papers are clustered and distributed, with each topic in a distinct color and outliers in gray. Note that clustering was performed in the reduced five-dimensional space, whereas the figure projects the embeddings onto two dimensions for display; it shows which papers group together more reliably than the distances between them. Dense clusters can be read as specialized areas and dispersed ones as multidisciplinary.

For instance, Topic 0 (Science Teacher Professionalism and Curriculum Implementation), the largest topic, shows the widest distribution. This may relate to its c-TF-IDF scores (see Fig. 4), where not only the first word but also the second and third words have high scores. Although BERTopic identified Topic 0 as a single topic, it encompasses varied themes, including preservice teachers, science teachers, achievement standards, and science inquiry. In contrast, Topic 8 (STEAM Education) occupied a narrow space. Its first word has a much higher score than the rest, and the remaining words are closely related to STEAM.

**Fig. 5**

*Visualization of vector embeddings tagged with topic names.*

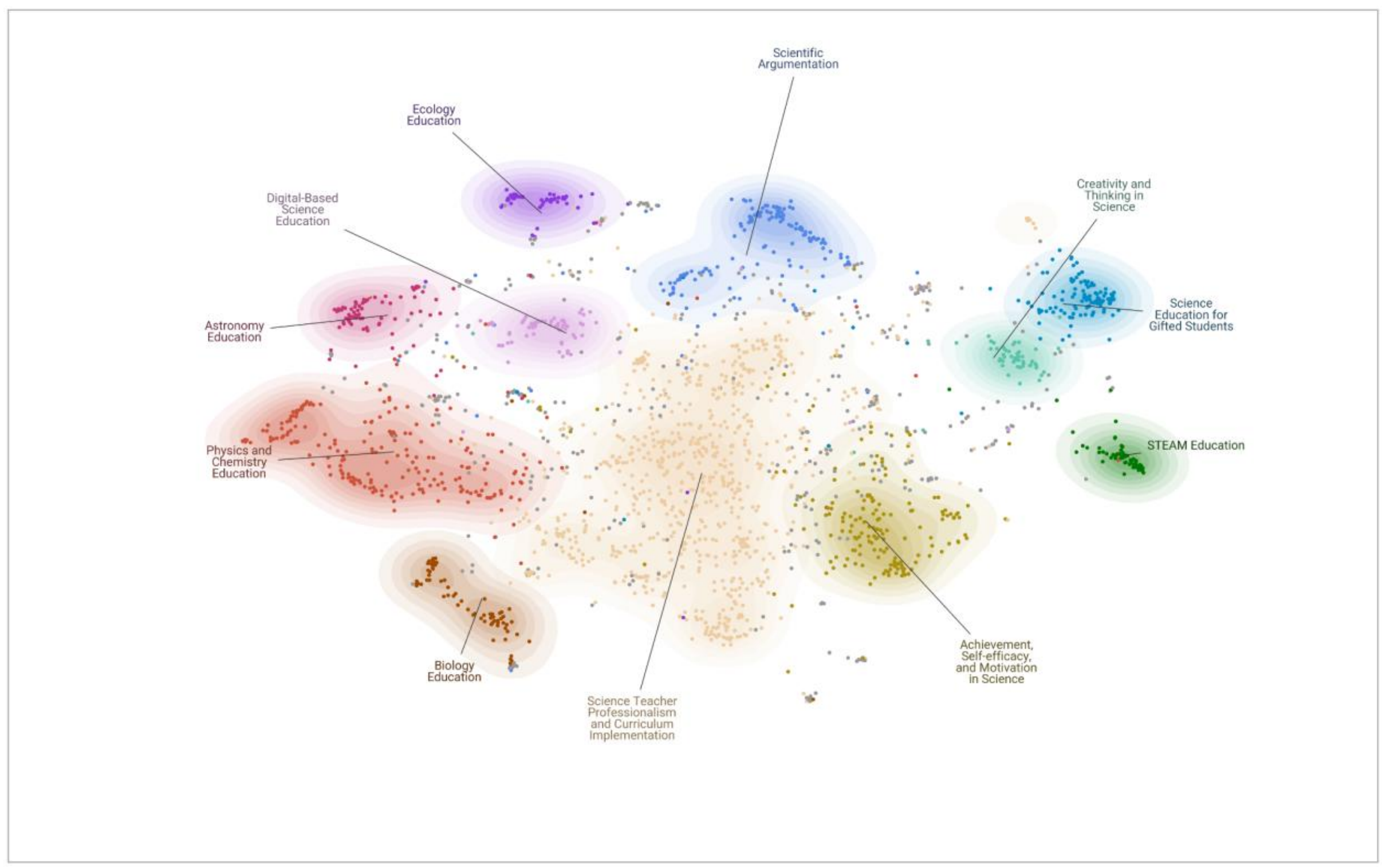

*Note.* Each axis is used only for spatial arrangement and does not imply direct semantic meaning.

## Topic Trends and Their Interpretations

Figure 6 illustrates topic trends from 2008 to 2025, with the y-axis showing the proportion of papers published per year for each topic. This proportion is the number of papers in a topic divided by the number of papers assigned to any topic that year. Outlier papers (Topic −1) were excluded, so the trend analysis is based on 1,873 papers. Figure 6a shows the overall topic trends, while Fig. 6b details each topic trend.

To examine whether topic trends relate to external conditions, we followed a two-step procedure. First,

after the topics were identified, we formed expectations about which of them should track external conditions. Based on each topic's representative words and papers, we identified topics closely related to policy, curriculum, or student demographics, and specified in advance of collecting the external data which external indicator each should correspond to—for example, gifted student enrollment for Topic 4 (Science Education for Gifted Students) and government STEAM projects for Topic 8 (STEAM Education). We then collected the corresponding data and tested the association. Because the indicators differ in kind, the form of the test differs accordingly: curriculum revision was treated as a period contrast, whereas enrollment and project counts were treated as continuous series. Second, we examined the temporal pattern of each topic, using the Mann-Kendall test to detect monotonic trends at $\alpha = .05$. This second step was applied to all topics, so that topics without a pre-specified indicator could also be checked for temporal structure.

Two points concern the data and the statistical procedure. Both external indicators are available only from 2012 (science-gifted enrollment from KEDI and project counts from KOSAC), so the analyses with external data cover 2012 to 2025 ($n = 14$). Given the small number of yearly observations, all $p$-values for the analyses with external data were obtained from permutation tests: pairing permutations for the Spearman correlations (Topics 4, 7, 8, and 9), and exact enumeration of all group assignments for the period contrast (Topic 0), which we complement with the Mann–Whitney $U$ test and report with Cliff's $\delta$, a rank-based effect size ranging from −1 to 1. Tests for the period contrast were one-sided, reflecting our prior expectation that publication activity increases following a curriculum announcement; all correlation tests were two-sided. Because publications may follow a funded project or a demographic change after some delay, we also examined one- to three-year lags for each pair. As none of the lagged correlations were significant or clearly stronger than the zero-lag correlation, we report the zero-lag results.

This procedure identified Topics 0, 4, 7, and 8 as candidates linked to specific external indicators. The Mann-Kendall test detected significant monotonic trends in Topics 4 ($\tau = -.60$, $p = .001$), 7 ($\tau = .66$, $p < .001$), and 9 ($\tau = -.39$, $p = .028$); the remaining topics showed no monotonic trend, although Topic 0 followed a periodic pattern that this test does not capture. We also expected Topic 7 (Digital-Based Science Education) to increase over time and to follow government digital-related projects; the increase was confirmed, but no association with project counts was detected. Topic 9 (Creativity and Thinking in Science) was not part of these pre-specified expectations. We examined it only after observing its resemblance to Topic 4, so its results should be read as exploratory.

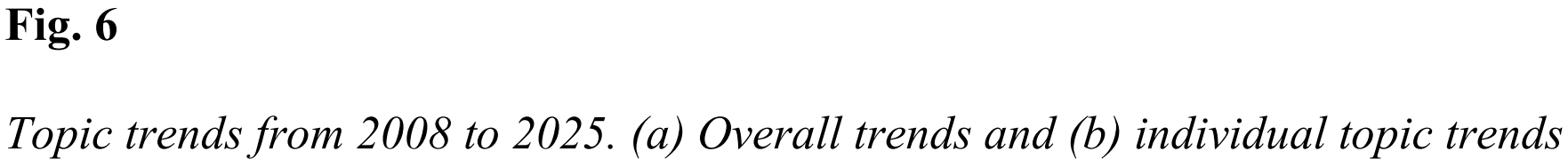

**Fig. 6**

*Topic trends from 2008 to 2025. (a) Overall trends and (b) individual topic trends*

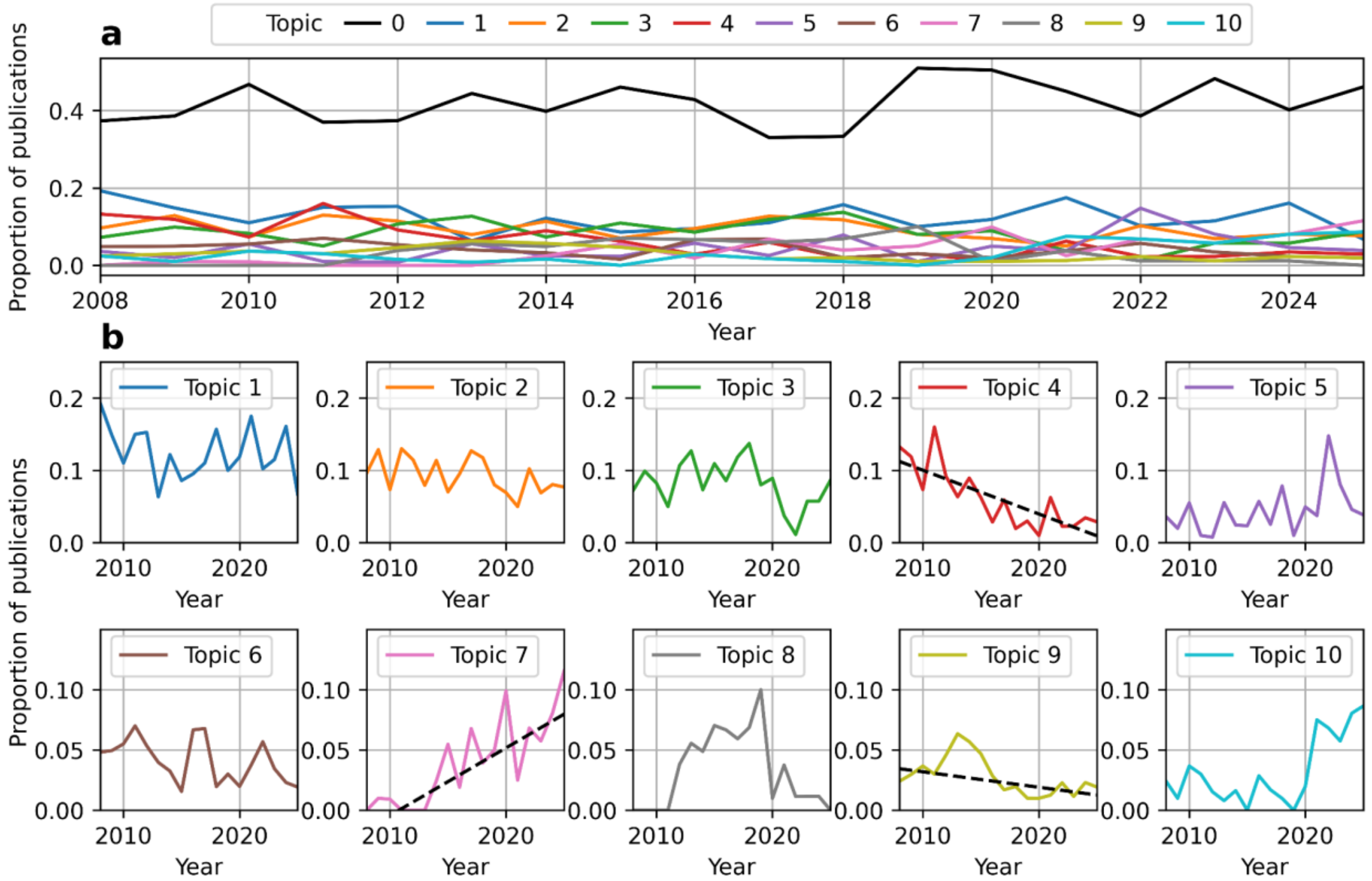

*Note.* Topic 0 is omitted from Fig. 6b; its trend appears in Fig. 6a. Topic numbers correspond to those in Table 1.

***Topics with Time-Series Trends***

Topic 0 (Science Teacher Professionalism and Curriculum Implementation) shows a periodic pattern corresponding to the national curriculum revision cycle, as shown in Fig. 7a. This topic concerns not only teacher expertise but also curriculum research, as indicated by representative words such as “achievement standards,” which are defined in the national curriculum itself (Ministry of Education, 2015, 2022). Previous studies have suggested that curriculum revision cycles influence research trends (Chang & Na, 2022; Kim et al., 2015), but these discussions have remained qualitative. We quantified this pattern by comparing annual changes in the topic’s proportion of publications. Following the announcement schedules of the 2009, 2015, and 2022 revised curricula, the year-over-year changes for 2012, 2013, 2018, 2019, 2020, and 2025 were classified as introduction-period changes, and the remaining eight as subsequent-period changes. The mean annual change was positive during the introduction periods ($M = 5.1$ %p, $n = 6$) and negative in the subsequent periods ($M = -2.7$ %p, $n = 8$). An exact, one-sided permutation test of this difference ($\Delta M = 7.8$ %p) yielded $p = .031$, and a one-sided exact Mann–Whitney *U* test yielded a consistent result ($U = 39$, $p = .030$, Cliff's $\delta = .63$). This periodicity can be understood as follows. Preliminary studies are conducted before national curriculum revision. After the official announcement, curriculum analysis research begins, including analyses of achievement standards and assessment criteria. However, due to textbook development, there is a 1–3 year gap between the announcement and actual school implementation. Considering the time required to conduct and publish research, the volume of research increases approximately two to three years after the announcement. Despite this cyclical pattern, a pronounced fluctuation occurred between 2021 and 2023, which may be associated with disruptions related to the COVID-19 pandemic.

Topic 4 (Science Education for Gifted Students) shows a decreasing trend, as shown in Fig. 7b. The proportion of publications on this topic has consistently declined, paralleling the decline in enrollment at science-gifted education centers (KEDI, n.d.). The Spearman correlation between the two series was significant ($\rho = .59$, $p = .030$), indicating a close temporal association between research output and gifted student enrollment. Beyond demographic decline, at least two additional factors may contribute. First, participation in gifted education programs no longer counts toward college admissions (D. Y. Kang, 2019). Second, a recent student preference for medical majors over the natural sciences and engineering is also reflected in this topic's dynamics (Seo et al., 2023). Topic 9 (Creativity and Thinking in Science) also shows a downward trend like that of Topic 4. This resemblance is consistent with the overlap between the two topics: they lie close to each other in the embedding space (see Fig. 5), and 11 of the 56 papers in Topic 9 directly concern science-gifted students. This overlap may help explain why Topic 9 follows a downward trend, as in Topic 4, and shows a strong association with gifted student enrollment ($\rho = .65$, $p = .014$).

Topic 8 (STEAM Education) emerged in 2011, driven by national efforts to raise students' interest in science and address the shortage of students pursuing STEM majors in Korea. This corresponds to the peak observed shortly after STEAM education was introduced, as shown in Fig. 7c. STEAM education research appears highly responsive to government-funded STEAM projects [external data from KOSAC (n.d.); we selected projects whose titles include "STEAM"], and the Spearman correlation between projects and publications was significant ($\rho = .75$, $p = .004$). This can be understood in terms of how STEAM education is positioned in Korea. STEAM education was introduced to address practical educational problems, which gives this topic a strong application orientation. At the same time, specialized STEAM degree programs are rare in Korea, so STEAM education has a weaker academic foundation than established disciplinary fields. This may explain why many studies on this topic focus on reporting the current status of STEAM education (see Supplementary Information 2), and why publication activity declined after policy attention and government support weakened.

Like STEAM education, digital-based science education has been proposed as a new educational agenda since 2011 (Joo et al., 2016). However, Topic 7 (Digital-Based Science Education) shows a weaker temporal association with the selected external data, and its Spearman correlation was not significant [see Fig. 7d; external data from KOSAC (n.d.); we selected projects whose titles include “Digital”]. This low correlation may relate to the following contexts ($\rho = .35$, $p = .209$). Digital devices are already widespread in everyday life and in schools, which makes their use in science lessons more feasible. Availability alone, however, does not ensure implementation (Fernandes et al., 2020; Yu & Chung, 2025); teachers’ beliefs and preparation shape how digital technologies are used (Yu & Chung, 2025). Still, when devices are readily accessible, teachers are more likely to use them, so classroom practice, and research on it, may grow with less dependence on government projects. By contrast, STEM/STEAM education requires cross-subject collaboration (Jho et al., 2016; Tytler et al., 2021), which is difficult to arrange in a subject-specialist teacher education system (Jho et al., 2016) and therefore depends more on external project support. This difference may help explain why publications on digital-based science education kept growing while those on STEAM education did not.

These analyses identify temporal correlations between topic prevalence and selected external data, but they do not establish causal inference. Accordingly, our interpretations should be read as possible indications of relative policy responsiveness, rather than as direct evidence about researchers’ motivations.

**Fig. 7**

*Trends for selected topics (colored lines) compared with external data (gray bars)*

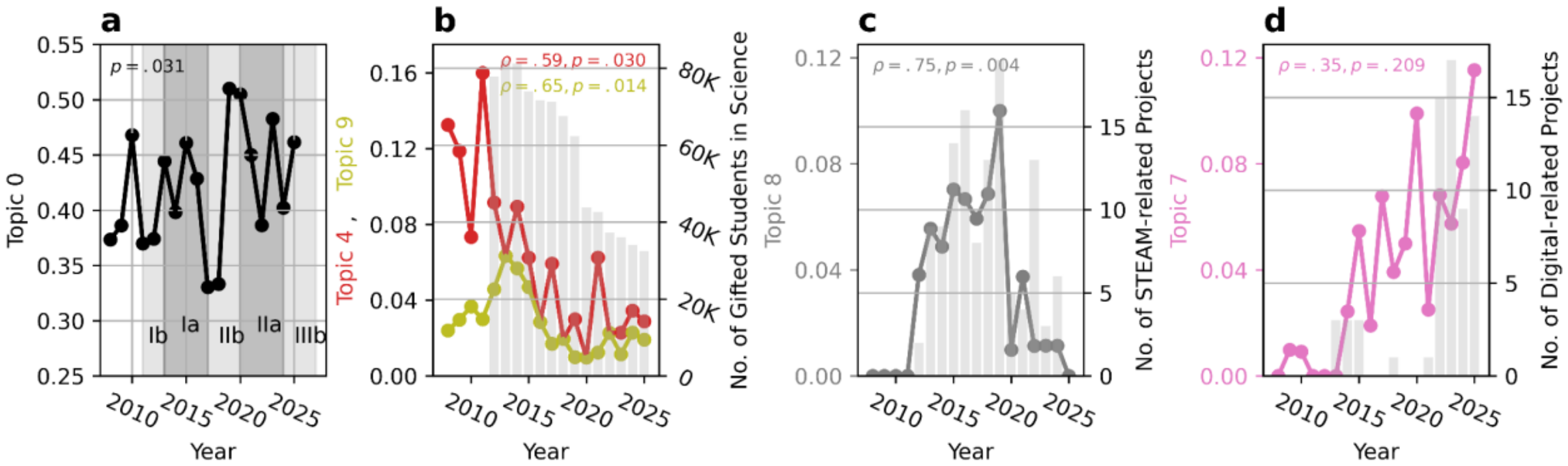

*Note.* The number showing the trend indicates the proportion of publications per year. (a) Light gray shades (with letter b) indicate the introduction periods of new revised curricula, and dark gray shades (with letter a) indicate the subsequent periods: (I) 2009 curriculum, (II) 2015 curriculum, and (III) 2022 curriculum. Gray bars represent (b) the number of gifted students in the science field, (c) the number of STEAM projects, and (d) the number of digital-related projects. Topic numbers correspond to those in Table 1.

***Topics without Time-Series Trends***

In contrast to the preceding topics, the remaining topics show a relatively consistent presence in the literature over time. They can be divided into two groups: subject-specific topics and student-related topics.

First, subject-specific topics (Topics 1, 5, 6, and 10) appear to be organized by internal academic agendas and the traditional discipline-based structure of Korean teacher education. These topics are grounded in physics, chemistry, biology, astronomy, and ecology. They are closely connected to the subject-based system of secondary teacher education in Korea. This suggests that disciplinary communities provide a solid foundation for these research agendas, making the topics less directly associated with sociopolitical conditions. This finding is consistent with Odden et al. (2021). Topic 10 (Ecology Education) may be a partial exception, as it shows a gradual increase in recent years. This may be related to growing attention to climate change. The Korean government designated "Ecological Transformation Education" as a major task (Ministry of Education, 2022) and incorporated sustainable development, the climate crisis, and ecological transformation into national educational goals, which may have contributed to the growth of related research after around 2022 (see Fig. 6b). However, this trend is too short to permit statistical verification, and additional data in the coming years may help clarify whether it is a longer-term pattern.

Second, student-related topics (Topics 2 and 3), except for research topics related to gifted students, are also less likely to be associated with external factors. Topic 2 (Achievement, Self-Efficacy, and Motivation in Science) seems to have been less affected, because studies in this area can draw on existing quantitative datasets. Topic 3 (Scientific Argumentation), by contrast, shows one notable valley during the COVID-19 period (see Fig. 6b), possibly because classroom-based studies on argumentation became more difficult to conduct at that time. Both topics are mainly grounded in established academic concerns in science education. Topic 3 deals with classroom discourse and students' reasoning; although studies of this topic have practical implications, they remain rooted in long-standing academic questions. Topic 2 addresses students' affective domain and career paths related to science, which connects academic concerns with broader societal interest in science participation. Their shared grounding in established academic agendas may explain why these topics form part of the stable core of science education research and show relatively constant publication trends.

## Conclusion

This study analyzed topics and temporal trends in selected Korean science education publications from 2008 to 2025 using BERTopic. Extending prior reviews centered on international journals, we examined not only which topics appear, but also how their trends relate to sociopolitical conditions and what distinguishes topics that track these conditions from those that do not.

We identified three groups of topics: (1) sociopolitical, (2) subject-specific, and (3) student-related topics. The first group showed distinct temporal trends or associations with external indicators, whereas the others did not. The sociopolitical topics suggest that when research is closely tied to policy, curriculum, or school-level needs, it is more likely to move with external conditions such as educational policy and demographic decline; STEAM education exemplifies this pattern most clearly. In contrast, topics with stronger disciplinary grounding showed more constant trends and weaker associations with sociopolitical conditions.

These findings point to a proposition that international researchers can examine elsewhere: a topic's responsiveness to external change depends less on the topic itself than on how it is anchored: whether in policy and practice or in an established disciplinary community. The Korean case supports this proposition in a clear form. Korean science education research shares many topics with international research (Chang & Na, 2022), but some differences appear. While STEM/STEAM education has continued to grow internationally (Chu, 2021; Georgiou et al., 2024), STEAM education in Korea rose and fell with government support. This suggests that, within Korea's highly centralized education system, application-oriented topics may respond quickly to policy demand but may require a stronger disciplinary foundation to be sustained. Practically, this implies that sustaining policy-driven applied research such as STEAM education requires more than government funding; its continuity also depends on a research–practice partnership grounded in shared conviction about the educational value of such work.

The same proposition can be tested directly across countries. Rather than comparing how often a topic appears in different national contexts, researchers could examine whether the same topic is anchored differently across countries, such as being driven by policy in one country but supported by established academic communities in another. The following example suggests that how a shared research agenda is handled can vary depending on local circumstances. Culture, equity, and gender have appeared as distinct topics in several international reviews (Lee et al., 2025; Lin et al., 2025; Odden et al., 2021), but they did not form separate topics in our analysis. Research on culture and equity appeared mainly within Topic 0 and, in the form of sustainability and climate-related concerns, within Topic 10. Gender-related research was somewhat more common, but it was concentrated on achievement differences and women's STEM career paths. This link between gender and STEM is not limited to Korea. Kahraman (2026) found a similar grouping in international journals, although "gender" was prominent enough to name the topic. However, our result should not be interpreted as an absence of these concerns. Cho et al. (2025) found that Korean elementary science textbooks rarely addressed human rights, discrimination, or equality, which they attributed to a curriculum focused on natural phenomena rather than social constructs. This finding provides one possible context for our result, although it does not directly explain the structure of the research literature. In addition, we excluded journals limited to specific subfields, so research on diverse learners may be published elsewhere outside of our sample. Therefore, our results suggest that culture, equity, and gender were present in Korean science education research, but were organized within broader topics rather than as distinct research agendas.

This study has limitations that suggest directions for future research. First, we analyzed three broad-scope journals and excluded those focused on specific subfields or school levels. As a result, some research areas may be underrepresented in our findings relative to Korean science education research as a whole. Second, although the external data helped identify temporal associations, they cannot provide direct evidence of causal relationships. In addition, each external indicator captures only part of the conditions surrounding a topic, so a weak association does not necessarily indicate that the topic is unrelated to external conditions. Future research could draw on broader sources and compare science education research across countries, showing not just how often a topic appears but how it is produced and sustained differently.